\documentclass[iop,apj]{emulateapj}

\usepackage{epsfig}
\usepackage{graphics,graphicx}
\usepackage{multirow}
\usepackage{gensymb}
\usepackage{amsmath,amssymb}
\usepackage{bm}
\usepackage{natbib}
\usepackage[outdir=./]{epstopdf}
\usepackage[breaklinks,colorlinks,urlcolor=blue,citecolor=blue,linkcolor=blue]{hyperref}
\usepackage{color}
\usepackage{chngcntr}

\def\h2{H$_{2}$}
\def\h1{H{\textsc i}}
\def\hii{H{\textsc i\textsc i}}
\def\12co{$^{12}$CO}
\def\13co{$^{13}$CO}
\def\nh3{NH$_{3}$}
\def\n2hp{N$_{2}$H$^{+}$}
\def\c2h{C$_{2}$H}
\def\cp{C$^{+}$}

\def\micron{$\mu$m}
\def\arcmin{$^\prime$}

\newcommand{\RNum}[1]{\uppercase\expandafter{\romannumeral #1\relax}}

\shorttitle{Catching the Birth of A Dark Molecular Cloud for the First Time}    
\shortauthors{Zuo et al.}

\begin{document}


\title{Catching the Birth of A Dark Molecular Cloud for the First Time} 

\author{Pei Zuo\altaffilmark{1,2}}
\affil{National Astronomical Observatories, Chinese Academy of Sciences, Datun Road, Chaoyang District, Beijing 100101, China} 
\affil{University of Chinese Academy of Sciences, Beijing 100049, China}
\email{peizuo@nao.cas.cn}

\author{Di Li\altaffilmark{1,2,3}}
\affil{National Astronomical Observatories, Chinese Academy of Sciences, Datun Road, Chaoyang District, Beijing 100101, China} 
\affil{University of Chinese Academy of Sciences, Beijing 100049, China}
\affil{Key Laboratory of Radio Astronomy, Chinese Academy of Sciences, Nanjing, 210008, China}

\author{J.~E.~G.~Peek\altaffilmark{4}}
\affil{Space Telescope Science Institute, 3700 San Martin Drive, Baltimore, MD 21218, USA}

\author{Qiang Chang\altaffilmark{5}}
\affil{Xinjiang Astronomical Observatory, Chinese Academy of Sciences, 150 Science 1-Street, Urumqi, Xinjiang 830011, China}

\author{Xia Zhang\altaffilmark{5}}
\affil{Xinjiang Astronomical Observatory, Chinese Academy of Sciences, 150 Science 1-Street, Urumqi, Xinjiang 830011, China}

\author{Nicholas Chapman\altaffilmark{6}}
\affil{CIERA - Northwestern University 2145 Sheridan Road Evanston, IL 60208, USA}

\author{Paul F.~Goldsmith\altaffilmark{7}}
\affil{Jet Propulsion Laboratory, California Institute of Technology, Pasadena, CA 91109, USA}

\and 

\author{Zhi-Yu Zhang\altaffilmark{8,9}} 
\affil{Institute for Astronomy, University of Edinburgh, Royal Observatory, Blackford Hill, Edinburgh EH9 3HJ, UK}
\affil{ESO, Karl Schwarzschild Strasse 2, D-85748 Garching, Munich, Germany}

\altaffiltext{1}{National Astronomical Observatories, Chinese Academy of Sciences, Datun Road, Chaoyang District, Beijing 100012, China}

\hyphenpenalty=5000
\tolerance=1000

\begin{abstract}

The majority of hydrogen in the interstellar medium (ISM) is in atomic form. The transition from atoms to molecules and, in particular, the formation of the H$_2$ molecule, is a key step in cosmic structure formation en route to stars. Quantifying H$_2$ formation in space is difficult, due to the confusion in the emission of atomic hydrogen (\h1) and the lack of a H$_2$ signal from the cold ISM. Here we present the discovery of a rare, isolated dark cloud currently undergoing H$_2$ formation, as evidenced by a prominent ``ring" of \h1\ self-absorption.Through a combined analysis of \h1\ narrow self-absorption, CO emission, dust emission, and extinction, we directly measured, for the first time, the [\h1]/[H$_{2}$] abundance varying from 2\% to 0.2\%, within one region. These measured \h1\ abundances are orders of magnitude higher than usually assumed initial conditions for protoplanetary disk models. None of the fast cloud formation model could produce such low atomic hydrogen abundance. We derived a cloud formation time scale of 6$\times$10$^6$ years, consistent with the global Galactic star formation rate, and favoring the classical star formation picture over fast star formation models. Our measurements also help constrain the H$_{2}$ formation rate, under various ISM conditions. 
    
\end{abstract}


\section{Introduction}

The formation of the H$_2$ molecules is a key process in converting atoms into molecules in the interstellar medium (ISM) and is the critical step in the formation of stars. \citet{1946Spitzer} suggested that H$_{2}$ formation occurs on dust grains, rather than in the gas phase, because the collision frequency is too low to provide sufficient three-body collisions in molecular clouds \citep{2002Kaiser}. A widely accepted description of an efficient \h1$-$H$_2$ transformation process in clouds is provided by \citet{1971Hollenbach}. Efficient H$_{2}$ formation can persist to high temperatures because chemisorbed atoms can stay on the grain surface while maintaining quantum mobility. Such configuration is conducive to hydrogen recombination (H$_{2}$ formation) until the dust grain surface temperature reaches $\sim$500~K \citep{2004Cazaux}. Laboratory experiments demonstrated that H$_{2}$ formation could also occur on carbon surfaces with a high recombination efficiency over a broad range of temperature \citep{2005Vidali}. The theoretical work in \citet{2005Cazaux} explained the high efficiency under low temperatures in both quiescent diffuse and dense clouds in terms of physisorption forces and thermally activated diffusion. Since the timescale of star formation is tightly constrained by the timescale of molecular gas formation, it is crucial to have a self-consistent understanding of the global star formation efficiency and star formation rate in galaxies. The timescale of star formation is controversial. The canonical model \citep{1977Shu} favored it to be around 10~Myr, while \citet{2004Bergin} and \citet{2000Elmegreen}, hypothesized that star formation occurs on a turbulent crossing scale that corresponds to about only 1~Myr on parsec scales \citep{2007Glover}. 

Direct measurement of H$_{2}$ formation is difficult because \h1 is hard to isolate in the Galaxy and H$_{2}$ lacks a permanent dipole moment, making it rare to have both measurements of \h1\ and H$_2$ gas for the same regions of the Galaxy. Key physical parameters need largely to be assumed in calculating cloud formation. Molecular hydrogen can be observed through absorption if it is against a good X-ray and/or ultraviolet background source (e.g.\ \citet{2001Sembach}). Since H$_2$ itself does not emit at the temperature of molecular clouds, these regions are generally traced emission from the lower rotational transitions of CO. However, subthermal excitation in diffuse gas and depletion in dense gas make CO unsatisfactory to trace H$_2$ \citep{1999Kramer}. It is also difficult to probe the cloud formation timescale through astrochemistry based on molecular species other than H$_2$ and \h1, such as N$_2$H$^+$ and deuterated species \citep{2002Caselli}, because these tracers comprise only a tiny fraction of cloud mass. The foremost chemical reaction in molecular formation is the formation of molecular hydrogen. It is thus paramount for us to obtain direct measurement of the H$_2$ formation timescale in order to better understand cloud formation timescale, the star formation efficiency and the star formation rate. 

We have developed an observational method called \h1 Narrow Self-Absorption (\h1NSA;  \citep{2003Li,2005Goldsmith}), which is capable of measuring atomic hydrogen abundance in dense clouds. Similar to the well-known \h1\ Self-Absorption (\h1SA \citep{1955Heeschen}) phenomenon in the cold neutral medium (CNM), \h1NSA differs in a significant aspect, namely, the thermalization of \h1 at low temperatures through collisions with H$_{2}$. Due to efficient cooling by molecular emission lines, especially the rotational transitions of CO, molecular clouds claim most of the lowest temperatures found in the Milky Way except for a few rare cases, such as those discussed by \citet{2001Knee} and \citet{2011Peekb}. In contrast, lacking molecular cooling, \h1SA is likely a result of temperature fluctuations in the CNM \citep{2000Gibson}.  Systematic surveys have shown that \h1NSA is associated with molecular clouds \citep{2010Krco}, and thus provides a direct  measurement of cold \h1 in dark clouds.

The left image in Figure~\ref{fig:model} gives a schematic diagram of the transition from CNM to dense molecular gas. In the warm ionized medium (WIM) and \hii\ regions, hydrogen is largely in the form of H$^+$. For higher densities and lower UV intensities, H$^{+}$ combines with electrons through radiative recombination and forms \h1. With increasing \h1 volume density, hydrogen atoms start to form molecular hydrogen on the surface of dust grains. In diffuse molecular clouds, even when most of the H is in the molecular form, gas-phase carbon can still be mainly \cp. After combining with H$_{2}$ through radiative association, \cp\ becomes CH$_{2}$$^{+}$; CH$_{2}$$^{+}$ reacts rapidly with electrons to form CH; then CH reacts with O to produce CO. \cp\ can also react with electrons to form CI through dielectronic recombination. CI, along with OH, lead the formation of CO through neutral$-$neutral reactions in colder (10\,K\,$<$\,$T$\,$<$\,100\,K) dark clouds \citep{2007Woodall}. Other molecules form following CO in dense cores. In particular,  N$_{2}$H$^{+}$, NH$_{3}$, and cations like H$_{2}$D$^{+}$ become significant and have been used  to study the cloud core formation time scale \citep{2014Brunken}.

Isolated, cold dark clouds are the ideal laboratory for studying H$_{2}$ formation in the Milky Way.In a previous study, B227, CB45, and L1574, were identified as isolated dark clouds, and coincidently aligned in a linear configuration from north to south spanning about two degrees \citep{1978Martin}. There are no nearby UV sources such as massive stars or \hii\ regions. 

In this paper, we present a combined analysis of \h1NSA, CO emission, dust emission, and extinction for a rare, isolated dark cloud, currently undergoing H$_2$ formation. The [\h1]/[H$_2$] abundance varies from 2\% to 0.2\%, within one region and the formation timescale was derived as $\sim$6~Myrs, which is consistent with both an analytical model and a numerical chemistry model.  

\section{Observations and Data Reduction}

To reconstruct the chemical state and the evolutionary history of isolated dark clouds, we mapped \h1, $^{13}$CO $J=1-0$ emission, and dust continuum emission in B227, CB45, and L1574, which were identified as \h1NSA sources by \citet{2005Goldsmith}. In 2012 May and November, the sources were observed in the 1420~MHz transition of \h1 using the Arecibo $L$-band Feed Array (ALFA) on the 305 m radio telescope. We implemented the observations in the ``total power" mode because of a clear ``off" position that was difficult to find in the Galaxy. The nominal system temperature of 30 K has contributions from the system temperature and the \h1\ emission. In addition, the temperature of the \h1\ emission from standard CNM is around 70~K \citep{2003Heiles}. Hence, we consider an equivalent system temperature of 100 K for evaluating the observing sensitivity. To accomplish the mapping for the sources that extend $1^{\circ}\times2^{\circ}$, the ``leap frog'' drift scan was adopted \citep{2007Minchin}. Spectra were recorded by GALSPECT, the Galactic ALFA spectrometer. The resolution is 0.18\,km\,s$^{-1}$ per channel with 7,679 channels. Using the Spectral and Photometric Imaging REceiver (SPIRE) on board the {\it Herschel Space Observatory}, the sources (B227, L1574, and CB45) were observed at 250, 350, and 500\,$\mu$m on 2011 September 11 within a {\it Herschel} Open Time (OT) 1 Program (OT1\_dli\_2). In the fast mapping mode having a nominal scan speed of 30$''$\,s$^{-1}$, two 9$'$ legs were used at each band for mapping each of the sources covered by the scanned area of $30' \times 30'$. \13co\ data were taken with the Five College Radio Astronomy Observatory (FCRAO) and adapted from \citet{2005Goldsmith}. 

Our SPIRE data were processed by using the software package HIPE\footnote{http:// http://herschel.esac.esa.int/hipe/; version 13.0.0}. The original image at each band was produced based on a zero-median brightness. To calibrate the offsets between bands, the \textsc{zeroPointCorrection} subroutine was used, which implements a cross-calibration with the {\it Planck} HFI-545 and HFI-857 images and a color correction HFI to SPIRE wavebands assuming a gray body function with fixed beta. We used the subroutine \textsc{Baseline Removal and Destriper} to remove the baseline with correcting the relative gain of the bolometer. \textsc{Photometer Map Merging} was then used to merge the images of the bands. The Arecibo \h1 data were processed following the procedure of the GALFA-\h1 Standard Reduction \citep{2011Peeka}. It is designed for the ALFA in order to take the time-ordered data (TOD) that come out of GALSPECT, over a single region and turn it into a calibrated, gridded spectral (PPV) data cube. The resulting data also showed stripes along the scan direction, we destriped the data by running the procedure to use the places of the data cross each other to self-calibrate the gains of each beam. We obtained Two Micron All Sky Survey (2MASS) extinction data based on the $J$, $H$, and $K_{s}$ bands of the 2MASS data. \13co data were taken with FCRAO and adapted from \cite{2005Goldsmith}. 
We regrided the pixel onto a 1\arcmin\ grid in order to match the \h1 data.

\section{Dust, Cold \h1\ and $^{13}$CO}
Our new continuous, Nyquist-sampled \h1\ map complemented by \13co\ and dust images enables us to identify a striking \h1NSA ring and directly measure the variation of \h1\ abundance in these transition clouds (Figure~\ref{fig:distribution}). 

\subsection{Dust Temperature and Column Density}
For each image pixel, an SED was extracted and fitted with a single-temperature modified blackbody of the form
\begin{equation}
    S_{\nu}=\Omega{B_\nu(\nu,\;T_\mathrm{d})}\kappa_\nu\mu{m_{\mathrm{H}}}N_{\mathrm{H_{2}}},
\label{eq:1}
\end{equation}
where $\Omega$ is the solid angle of the emitting element, $B_{\nu}$ is the blackbody emission from the dust at temperature $T_{\mathrm{d}}$, 
$\kappa_{\nu}$ is the dust mass absorption coefficient, $\mu=2.33$ is the particle mass per hydrogen molecule, 
$m_{\mathrm{H}}$ is the mass of hydrogen atom, and $N_{\mathrm{H_{2}}}$ is the column density of hydrogen molecules, 
and we assumed the gas-to-dust ratio is 100. In Equation (\ref{eq:1}),
\begin{equation}
    B_{\nu}(\nu,\;T_{\mathrm{d}})=\frac{2h\nu^{3}}{c^{2}}\frac{1}{e^{(h\nu/kT_{\mathrm{d}})}-1}
\end{equation}
is the Planck function and
\begin{equation}
    \kappa_{\nu}=\kappa_{230}\big(\frac{\nu}{230\;\mathrm{GHz}}\big)^{\beta},
\end{equation}
(where $\kappa_{230}$ = 0.009$\;\mathrm{cm^{2}\;g^{-1}}$) is the emissivity of the dust grains at 230 GHz.

We used 2MASS extinction data to eliminate the column density of hydrogen atoms as a parameter for fitting the dust temperature $T_{\mathrm{d}}$ and emissivity spectral index $\beta$. The hydrogen column density $N_{\mathrm{H}}$ can be estimated from the optical extinction \citep{2009Guver},
\begin{equation}
\frac{N_{\mathrm{H}}}{\mathrm{cm^{-2}}}=2.2\times10^{21}\left(\frac{A_{\mathrm{v}}}{\mathrm{mag}}\right).
\end{equation}
As the atomic hydrogen fractional abundance is small, we take $N_{\mathrm{H{\textsc i}}}=0$ and obtain
\begin{equation}
N_{\mathrm{H}}=2N_{\mathrm{H_{2}}}.
\end{equation}
The column density of hydrogen molecules $N_{\mathrm{H_{2}}}$ is related to visual extinction $A_{\mathrm{v}}$,
\begin{equation}
\frac{N_{\mathrm{H_{2}}}}{\mathrm{cm^{-2}}}=1.1\times10^{21}\left(\frac{A_{\mathrm{v}}}{\mathrm{mag}}\right).
\end{equation}
Combined {\it Herschel} SPIRE 250, 350 and 500 \micron\ data with {2MASS} extinction data, the dust temperature $T_{\mathrm{d}}$, column density of hydrogen molecules $N_{\mathrm{H_{2}}}$ and emissivity spectral index $\beta$ can be fitted simultaneously by using Equation (\ref{eq:1}).

\subsection{Column Density of Cold \h1}
The new Arecibo \h1\ images provide us with the overall \h1NSA distribution in these sources. In order to extract the absorption component, we subtracted a relatively uniform background by averaging the \h1 emission values from surrounding points of the clouds, which has been done in a fashion similar to that described by \citet{2001Knee}. The method for solving for the optical depth of the absorbing gas is described from Eq.~12 in \citet{2003Li}, 

\begin{eqnarray}
\tau_{0}=ln[\frac{pT_{\mathrm{\h1}}+(T_{\mathrm{c}}-T_{\mathrm{x}})(1-\tau_{\mathrm{f}})}{pT_{\mathrm{\h1}}+(T_{\mathrm{c}}-T_{\mathrm{x}})(1-\tau_{\mathrm{f}})-T_{\mathrm{ab}}}],
\end{eqnarray}
where $\tau_{0}$ is the peak optical depth of the absorption feature at the centroid velocity, $\tau_{\mathrm{f}}$ is the foreground \h1 optical depth, taken to be 0.1, which is the typical value of the foreground \h1\ and only results in 0.4\% uncertainty of optical depth of \h1. $T_{\mathrm{\h1}}$ is the quantity obtained from the spectrum by fitting a polynomial to the portion of the spectrum without the absorption, $T_{\mathrm{c}}$ is continuum temperature, taken to be 3.5~K, including the cosmic background and Galactic continuum emission. $T_{\mathrm{x}}$ is the excitation temperature of \h1 in the dark cloud, which could be replaced by the dust temperature, $T_{\mathrm{ab}}$ is the absorption temperature, which is referred to as the depth of absorption line, $p$ is the ratio of background \h1\ optical depth $\tau_{\mathrm{b}}$ to optical depth of the uniformed Galactic atomic gas $\tau_{\mathrm{h}}$ ($\tau_{\mathrm{b}}=p\tau_{\mathrm{h}}$), $\tau_{\mathrm{h}}$ is the total of the foreground \h1\ optical depth and the background \h1\ optical depth ($\tau_{\mathrm{h}}=\tau_{\mathrm{f}}+\tau_{\mathrm{b}}$), and $p$ can be calculated from a model of the local \h1\ distribution and is generally in the range of 0.8-0.9. 

The column density of \h1 is given by
\begin{equation}
    \frac{N_{\mathrm{\h1}\mathrm{NSA}}}{\mathrm{cm^{-2}}}=1.95\times10^{18}{\tau_{0}}\frac{\Delta{V}}{\mathrm{km\;s^{-1}}}\frac{T_{\mathrm{k}}}{\mathrm{K}},
\end{equation}
where $\Delta$$V$ is the FWHM of the absorption line from a Gaussian fit, and $T_{\mathrm{k}}$ is the kinetic temperature. We used the dust temperature $T_{\mathrm{d}}$ derived from {\it Herschel} SPIRE maps as $T_{\mathrm{k}}$, which should be well coupled in such clouds \citep{2005Goldsmith}. The column density of cold \h1 had been obtained based on \h1NSA analysis. 

Since the true total column density measurement will be affected by the optically thin approximation, several approaches have been employed to evaluate this potential problem. \citet{2015Lee} used two different methods to estimate the correction factor $f$ for high optical depth and found that they are consistent, which is likely due to the relatively low optical depth and insignificant contribution from the diffuse radio continuum emission. As the clouds we study are located in the outer region of the Galaxy and the optical depths are small \citep{2002Kolpak}, the correction for the optical depth should not affect the \h1 column density significantly. The overall uncertainty is approximately 50\%. 

Since we cannot obtain the $T_\mathrm{k}$, we present the upper and lower limit \h1\ column density and \h1\ abundance by using temperature of 10 and 30 K. The upper and lower limits for the \h1\ column density for the three clouds are 7.50$\times10^{19}$cm$^{-2}$ and 2.31$\times10^{18}$cm$^{-2}$ respectively. The upper and lower limits for the \h1\ abundance are 0.09 and 0.001 respectively.

\subsection{\13co Column Density}
The central frequency of the \13co $J=1-0$ line $\nu$ is 110.2\,GHz. 

The column density of \13co in the upper level ($J=1$) can be expressed as 
\begin{equation}
N_{\mathrm{u, ^{13}CO}}=\frac{8{\pi}k{\nu}^{2}}{hc^{3}A_{\mathrm{ul}}}\int{T_{\mathrm{b}}}dv,
\end{equation}
where $k$ is Boltzmann's constant, $h$ is Planck's constant, $c$ is the speed of light, $A_{\mathrm{ul}}$ is the spontaneous decay rate from the upper level to the lower level, and $T_{\mathrm{b}}$ is the brightness temperature. 
We obtain
\begin{equation}
\left(\frac{N_{\mathrm{u, ^{13}CO}}}{\mathrm{cm^{-2}}}\right)=3.7\times10^{14}\int\left({\frac{T_{\mathrm{b}}}{\mathrm{K}}}\right)d\left(\frac{v}{\mathrm{km\;s^{-1}}}\right).
\end{equation}

The total \13co column density $N_{\mathrm{tot}}$ is related to the upper level column density $N_{\mathrm{u}}$ through

\begin{equation}
N_{\mathrm{tot, ^{13}CO}}=f_{\mathrm{u}}f_{\tau}f_{\mathrm{b}}N_{\mathrm{u, ^{13}CO}},
\end{equation}
where the level correction factor $f_{\mathrm{u}}$ can be calculated analytically under the assumption of local thermal equilibrium (LTE) as
\begin{equation}
f_{\mathrm{u}}=\frac{Q(T_{\mathrm{ex}})}{g_{\mathrm{u}}\mathrm{exp}\left(-\frac{h\nu}{kT_{\mathrm{ex}}}\right)},
\end{equation}
where $g_{\mathrm{u}}$ is the statistical weight of the upper level. $T_{\mathrm{ex}}$ is the excitation temperature and $Q(T_{\mathrm{ex}})$ = $kT_{\mathrm{ex}}/B_{\mathrm{e}}$ is the LTE partition function, where $B_{\mathrm{e}}$ is the rotational constant \citep{2005Tennyson}. The partition function can be expressed as $Q(T_{\mathrm{ex}})$ $\approx$ $T_{\mathrm{ex}}/ 2.76\;\mathrm{K}$. The correction factor for opacity $f_{\tau}$ is defined as
\begin{equation}
f_{\tau}=\frac{\int{\tau_{13}}dv}{\int{(1-e^{-\tau_{13}})dv}} ,
\end{equation}
and the correction for the background $f_{\mathrm{b}}$ can be expressed as
\begin{equation}
f_{\mathrm{b}}=\left[1-\frac{e^{\frac{h\nu}{kT_{\mathrm{ex}}}}-1}{e^{\frac{h\nu}{kT_{\mathrm{bg}}}}-1}\right]^{-1},
\end{equation}
where $\tau_{13}$ is the opacity of the \13co transition and $T_{\mathrm{bg}}$ is the background temperature, which is assumed to be 2.7~K. The \13co opacity is estimated under the assumption of being optically thin, for which $f_{\tau}=1$. The excitation temperature $T_{\mathrm{ex}}$ can be replaced by the dust temperature $T_{\mathrm{d}}$. The maximum $^{13}$CO column densities obtained using this analysis for the three sources B227, L1574, and CB45 are $4.1\times 10^{15}$ cm$^{-2}$, $4.9 \times 10^{15}$ cm$^{-2}$, and $7.3 \times 10^{15}$ cm$^{-2}$. We then derived the \13co\ abundance through column densities of dust and \13co with consideration of other chemical effects, like \13co depletion, and all of them are likely not dominant for our case. 

Based on the dust data, the central volume densities of the three clouds, B227, L1574, and CB45 are 1600 cm$^{-3}$, 2200 cm$^{-3}$, and 1800 cm$^{-3}$, assuming cloud sizes of 0.8~pc, 1.0~pc, and 1.5~pc respectively. The volume density is much greater than the \13co\ critical density of 1266 cm$^{-3}$, confirming the validation of the LTE assumption.

\subsection{Abundances}

\begin{figure*}[ht!]
    \begin{center}
           \includegraphics[angle = 0, trim =0cm 0cm 0cm 0cm,width=18cm]{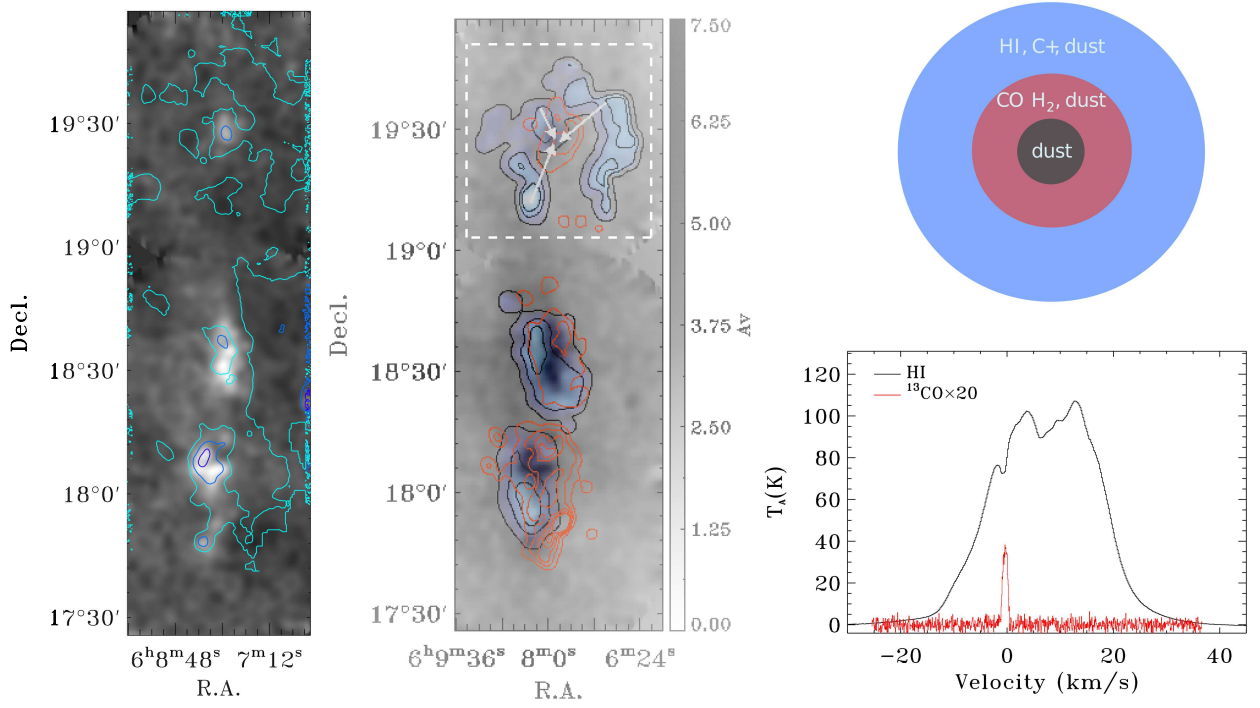}           
       \end{center}
       \caption{Temperature distribution, abundance distribution, onion-like model, and \h1NSA feature. 
Left: the 2MASS extinction overlaid with the dust temperature. The contours from light blue to dark blue are 12~K, 13~K, and 14~K respectively. 
Middle: the 2MASS extinction overlaid with the ratio of [\h1]/[H$_2$] column densities (black contour and blue shadow), and [\13co]/[H$_{2}$] column density (red) respectively. The black contours are at 10\%, 40\%, and 70\% of maximum value of 2.1$\times$10$^{-2}$. The red contours are at 10\%, 30\%, 50\%, 70\%, and 90\% of maximum value. The maximum value is 2.1$\times10^{-6}$. The white arrows show the direction of the \h1-H$_{2}$ transition of B227. 
Right top: an onion-like model of evolving molecular cloud, which is a 2D version of the left image, corresponding to the source B227. 
Right bottom: \h1NSA feature of the source B227 at velocity $\simeq-1$ km s$^{-1}$, with corresponding \13co\ emission, which is from the center of B227. 
}
      \label{fig:distribution}
 \end{figure*}

The abundances of \h1NSA and \13co\ have been derived through a combined analysis with dust emission and extinction. We regarded the ratio between \h1\ column density $N_{\mathrm{\h1}}$ and H$_2$ column density $N_{\mathrm{H_2}}$ as \h1\ abundance ([\h1]/[H$_2$]). The \13co\ abundance is derived from the ratio between \13co\ column density $N_{\mathrm{^{13}CO}}$ and H$_2$ column density $N_{\mathrm{H_2}}$ ([\13co]/[H$_2$]). We identified a striking ``ring" of enhanced \h1\ abundance, which is the first time such a structure has been seen in a molecular cloud. It closely resembles the ``onion" shell description of a forming molecular cloud (Figure~\ref{fig:distribution}). The displacement between peak abundance positions of \h1, \13co, and dust indicates the ongoing H$_2$ formation. Particularly, we find an orderly spatial variation of \h1\ abundance between 2\% and 0.2\%. The distribution of \h1\ abundance measured from \h1NSA is also consistent with the inner cloud ``core" being chemically more evolved, due to higher volume densities present there. 

\section{Model}

\subsection{H$_2$ Formation Model}

H$_{2}$ formation occurs on the dust grain with a formation rate \citep{1971Hollenbach}
\begin{equation}
R_{\mathrm{H_{2}}}=\frac{1}{2}S_{\mathrm{\h1}}\epsilon_{\mathrm{H_{2}}}n_{\mathrm{\h1}}\langle{v_{\mathrm{\h1}}}\rangle\sigma_{\mathrm{gr}}n_{\mathrm{gr}},
\end{equation}
where $R_{\mathrm{H_2}}$ has units of $\mathrm{cm^{-3}\,s^{-1}}$, $S_{\mathrm{\h1}}$ is the sticking efficiency, $\epsilon_{\mathrm{H_{2}}}$ is the recombination efficiency, $n_{\mathrm{gr}}\, \mathrm{(cm^{-3})}$ is the atomic hydrogen density in the gas, $\langle{v_{\mathrm{\h1}}}\rangle\,\mathrm{(cm\;s^{-1})}$ is the mean velocity of atoms, $\sigma_{\mathrm{gr}}\,\mathrm{(cm^2)}$ is the cross section of grains, and $n_{\mathrm{gr}}\,\mathrm{(cm^{-3})}$ is the number density of grains. 

The values and derivations of the parameters are taken from \cite{2005Goldsmith}. The simplification of the H$_2$ formation rate can be expressed as
\begin{equation}
R_{\mathrm{H_2}}=k_{\mathrm{H_2}}n_{\mathrm{gas}}n_{\mathrm{\h1}}.
\end{equation}
We denote the $\mathrm{H_2}$ formation rate coefficient $k_{\mathrm{H_2}}$ as $k'=1.2^{-17}$\,cm$^{3}$\,s$^{-1}$ after taking the effect of grain size distribution into account \citep{2005Goldsmith}. 

The formation rate coefficient $k'$ is related to the grain size, namely the cross-section area of dust grains, which is connected to the thermal balance and the estimate of dust mass. Based on the study of grain size distribution from \citet{1977Mathis}, the reasonable upper and lower limits of radii, $a_\mathrm{max}=10000$\AA\ and $a_\mathrm{min}=25$\AA, which enter as  ($a_\mathrm{max}a_\mathrm{min}$)$^{-0.5}$, are adopted for evaluating the effect of the grain size distribution on the formation rate coefficient. In comparison with the standard grain radius 1700\AA, the formation rate coefficient is increased by a factor of 3.4. Combining the grain size distribution and other factors, the uncertainty of H$_2$ formation rate coefficient may differ from its nominal value by a factor of 5. Although the nominal value of $k'$ we take has large uncertainty, it is well beyond the goal of this paper to consider the overall context. Therefore, we are utilizing the most common value for the cross-section area of the dust. 

The density of gas $n_{\mathrm{gas}}$ can be regarded as total density $n_0$ as $n_0\propto n_{\mathrm{gas}}$. We derived the volume proton density of the clouds through the dust data. For the volume proton density, we derived the mean values from outer to inner of the cloud by assuming the cloud is spherical. It ranges from $10^{2.9}$\,cm$^{-3}$ to $10^{3.2}$\,cm$^{-3}$ approximately (Figure~\ref{fig:model}). The time dependence of the molecular hydrogen can be express as
\begin{equation}
\frac{dn_{\mathrm{H_{2}}}}{dt}=k'n_{\mathrm{\h1}}n_{0}-\zeta_{\mathrm{H_{2}}}n_{\mathrm{H_{2}}},
\end{equation}
where $\zeta_{\mathrm{H_{2}}}$ is the cosmic-ray ionization rate. We defined the fractional abundance of atomic and molecular hydrogen, $x_{\mathrm{\h1}}$ and $x_{\mathrm{H_{2}}}$, as the density of species $n_{\mathrm{\h1}}$ and $n_{\mathrm{H_2}}$ divided by the total proton density $n_0$, then we find 
\begin{equation}
\frac{dx_{\mathrm{H_{2}}}}{dt}=k'x_{\mathrm{\h1}}n_{0}-\zeta_{\mathrm{H_{2}}}x_{\mathrm{H_{2}}}. 
\end{equation}
Substituting $n_0 = n_{\mathrm{\h1}}+n_{\mathrm{H_2}}$, we can rewrite this as 
\begin{equation}
\frac{dx_{\mathrm{H_{2}}}}{dt}=k'n_{0}-(2k'n_0+\zeta_{\mathrm{H_{2}}})x_{\mathrm{H_{2}}}. 
\end{equation}
The time dependence of the fractional abundance of molecular hydrogen is then given by 
\begin {equation}
x_{\mathrm{H_2}}(t) = \frac{k'n_0}{2k'n_0+\zeta_{\mathrm{H_2}}}[1-\mathrm{exp}(\frac{-t}{\tau_{\mathrm{\h1}\rightarrow\mathrm{H_{2}}}})].
\end{equation}
The fractional abundance of atomic hydrogen is $1-2x_{\mathrm{H_2}}$ and is given by
\begin{equation}
x_{\mathrm{\h1}}(t)=1-\frac{2k'n_0}{2k'n_0+\zeta_{\mathrm{H_2}}}[1-\mathrm{exp}(\frac{-t}{\tau_{\mathrm{\h1}\rightarrow\mathrm{H_{2}}}})].
\end{equation}
The timescale for \h1\ to H$_2$ conversion is given by 
\begin{equation}
\tau_{\mathrm{\h1}\rightarrow\mathrm{H_{2}}}=\frac{1}{2k'n_0+\zeta_{H_2}}.
\end{equation}

We derived the timescales of these isolated dark clouds to be {\color{blue}$\sim$6}~Myrs using \h1\ abundance, total gas density, and gas temperature obtained according to the time dependence model described in this section (Figure~\ref{fig:model}). The upper and lower limits of \h1\ abundances are plotted in Figure~\ref{fig:model} and the points at upper limits represent the lower ages of the clouds for the fixed volume densities.

\subsection{Chemical Model}

We also use a two-stage young dark clouds formation collapse model. We assume that translucent clouds are the intermediate stage when diffuse clouds become young dark clouds. In the first stage, when diffuse clouds become translucent clouds, we only consider the \h1 $\rightarrow$ H$_{2}$ transition process as discussed in the rest of the paper. In the second stage, we simulate a gas$-$grain reaction network to calculate the chemical evolution of species including \h1 and H$_{2}$ in collapsing cores. 

\begin{figure*}[ht!]
    \begin{center}
        \includegraphics[angle = 0, trim =0cm 0cm 0cm 0cm,width=18cm]{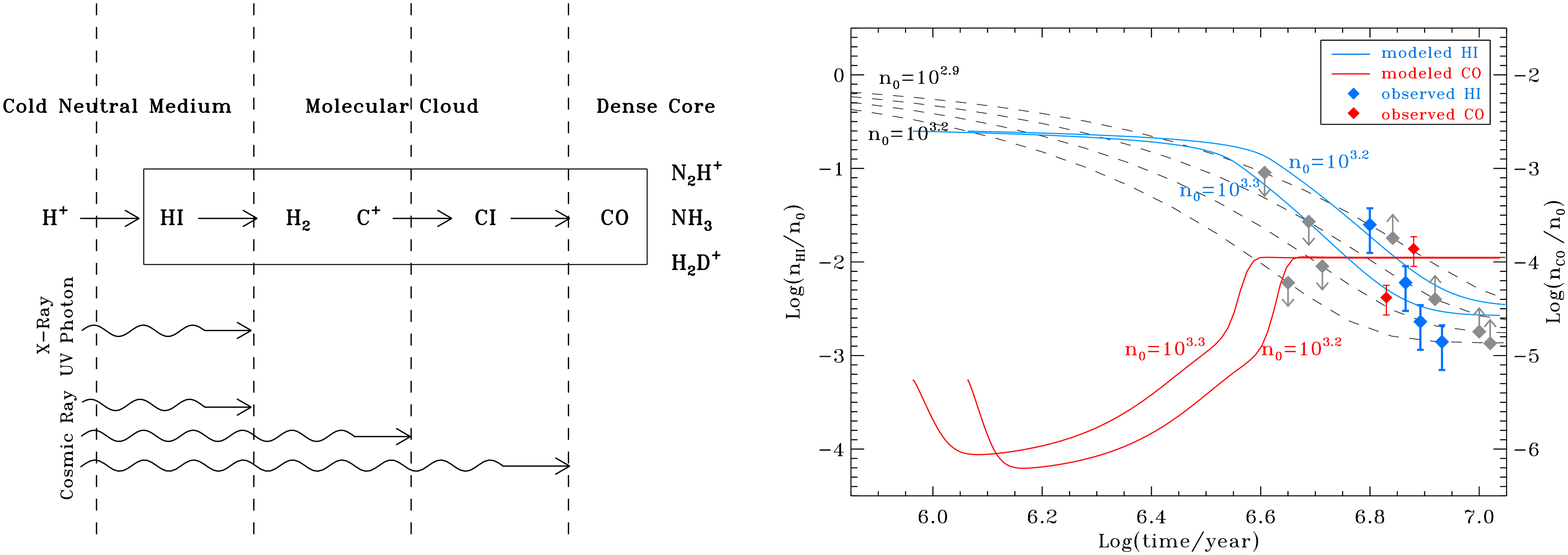}
                  
     \end{center}
              \caption{Chemical process and time-dependent model. 
Left: the chemical processes of diffuse CNM$-$molecular cloud$-$dense core. 
The processes in the square indicates the chemical evolution in our work, 
which is corresponding to the right plot. 
Right: time dependence of the fractional abundance of the atomic hydrogen and CO. 
The blue and red symbols represent the \h1\ and CO observations, respectively. 
The four \h1\ measurements are from the four points that are along the right top arrow of source B227 in Figure~1. 
The two CO measurements (from left to right) are the mean values from the outer envelop and inner core of the source B227, 
and the mean proton total densities are 10$^{2.9}$ cm$^{-3}$ and 10$^{3.0}$ cm$^{-3}$ respectively. 
The dashed lines represent the H$_2$ formation model for clouds at different total proton densities $n_0$. 
The blue and red lines represent the cloud evolution in a time-dependent model with the abundance of \h1 and CO. 
The gray diamonds with upward and downward arrows represent the lower and upper limits of the \h1\ abundances. 
}
      \label{fig:model}
 \end{figure*}

We use the Ohio State University (OSU) gas$-$grain code. The gas$-$grain chemical reaction network is explained in \cite{2011Hincelin} and \cite{2013Hincelin}. Moreover, the Kinetic Database for Astrochemistry (KIDA) database \citep{2012Wakelam} also has an electronic version of the network. We simulate the gas$-$grain chemistry in free-fall collapsing cores with initial hydrogen nucleus densities of $10^{2.3}$ and $10^{2.4}$ cm$^{-3}$ and evolve to $10^{3.2}$ and $10^{3.3}$ cm$^{-3}$ eventually. It takes $10^{7}$ years to collapse to form a dense core. During free-fall collapse, the rate at which the density increase is give by \cite{1978Spitzer}. The collapsing process is isothermal with a constant temperature of 10~K while the visual extinction varies from 1 to 4~mag. The cosmic-ray ionization rate is set to 5.2$\times10^{-17}$ s$^{-1}$. The size of dust grains is assumed to be uniform, with a radius of 0.1\,$\mu$m. The sticking coefficient is fixed to be 1. The elemental abundances are assumed to be ``low metal," as in \cite{2010Semenov}. In our model, species in the young dark clouds are formed by the chemical evolution of species in translucent clouds. Thus, we use the observed abundances of species in translucent clouds in HD 24534, HD 154368 and HD 210121 \citep{2004Sofia,2007Sonnentrucker,2009Weselak,2010Burgh} where the atomic H density is about a quarter of the total H nucleus density of the initial abundances in our calculation of the second stage. If a species is observed in more than one source, we take the average of the observed values in different sources. Table~\ref{tbl:chemi_model} shows the initial abundances used in our calculation. Through this time-dependent chemical model, we compared the timescales of cloud evolution traced by the abundance of \h1\ and CO with our observed abundances and calculated timescale. The observed \h1\ abundance and timescale of cloud formation is partially consistent with the chemical model. On the other hand, CO abundance is only marginally consistent with the chemical model, which already evolved into the steady state at the corresponding abundance (Figure~\ref{fig:model}). 

\begin{table}[ht!]
\caption{Initial abundances used in the chemical model\label{tbl:chemi_model}}
\begin{center}
\begin{tabular}{lcccc}
\hline
\hline
Element  & Abundance \\
\hline
C  & $4.34\times10^{-7}$
\\
H & $2.50\times10^{-1}$
\\
He  & $9.00\times10^{-2}$
\\
N  & $7.60\times10^{-5}$
\\
O  & $2.50\times10^{-4}$
\\
C$_{2}$  & $1.79\times10^{-8}$
\\
CH  & $1.59\times10^{-8}$
\\
CN  & $6.79\times10^{-9}$
\\
CO  & $5.45\times10^{-6}$
\\
H$_{2}$  & $3.75\times10^{-1}$
\\
NH  & $1.78\times10^{-9}$
\\
OH  & $4.36\times10^{-8}$
\\
C$^{+}$  & $1.14\times10^{-4}$
\\
Cl$^{+}$  & $1.00\times10^{-9}$
\\
Fe$^{+}$  & $3.00\times10^{-9}$
\\
Mg$^{+}$  & $7.00\times10^{-9}$
\\
Na$^{+}$  & $2,00\times10^{-9}$
\\
P$^{+}$  & $2.00\times10^{-10}$
\\
S$^{+}$  & $8.00\times10^{-8}$
\\
Si$^{+}$  & $8.00\times10^{-9}$ 
\\
CH$^{+}$  & $5.51\times10^{-9}$
\\
e  & $1.14\times10^{-4}$  
\\
\hline
\end{tabular}
\end{center}
\end{table}

\section{Discussion}

It has been rare to have both \h1\ and H$_2$ measured for the same regions in the Galaxy. \citet{2015Lee} used the integrated \h1\ emission flux in channels, the range of which was determined by correlation with 2MASS extinction. Such a priori requirements of \h1\ mimicking dust diminish the logical credence of \h1\ emission being tracing the atomic component within a molecular cloud. \citet{2014Fukui} found an optically thick \h1\ emission envelope around a molecular cloud. In this work, utilizing \h1NSA, we discovered a chemically young molecular cloud undergoing H$_2$ formation, revealed by a prominent ``ring" of self-absorption tracing atomic hydrogen mixed with molecules. We derived the formation timescale to be $\sim$6~Myr, consistent with both an analytical model and a numerical chemistry model. Our results could further test recent H$_{2}$ formation models in different contexts \citep{2014Sternberg,2017Bialy}. 

Our results can also help constrain galaxy evolution simulations. A key recent development is the implementation of the physics of molecular hydrogen and star formation prescription based on the local molecular hydrogen density rather than the total hydrogen density \citep{2012Zemp}. In \citet{2011Gnedin}, only two types of observations were able to calibrate the H$_2$ formation rate in different environments, UV absorption in diffuse regions and \h1NSA in dense clouds. This work further improves the accuracy of the \h1NSA measured H$_2$ formation rate.

Both the timescale and \h1\ abundance distribution of the clouds are inconsistent with fast H$_2$ formation \citep{2007Glover}, which relies on stochastic processes of localized high volume \h1\ density. Such a fast cloud formation model could produce a significant amount of molecular hydrogen, e.g.\ for cooling in a short time, but cannot convert the vast majority of the gas into molecular form, which is our key finding here. The fact that more than 99\% of the protons have been turned into H$_2$ requires millions of the years and comprehensive global H$_2$ formation, not just in pockets of locally enhanced condensations. The observed Mach number of the clouds is more than 20, which elevates the inconsistency between our observation and the fast cloud formation model. 

The total column densities found in these isolated dark clouds are consistent with commonly used initial conditions for protoplanetary disks. However, the measured \h1\ abundance is one to a few orders of magnitude higher than those assumed in such models (e.g.\ \citealt{2015Walsh}). In these models, the \h1\ abundance keeps dropping with time and impacts subsequent chemical reactions. We expect our measured value to affect future consideration of chemical evolution of disks. All the above statements indicate that the parameters of galactic H$_2$ formation play a crucial role in both galaxy evolution and star formation.

\section{Summary}

We mapped \h1, $^{13}$CO $J=1-0$ emission, and dust continuum emission of isolated dark clouds, B227, CB45, and L1574. The combined analysis of \h1NSA, CO emission, dust emission, and extinction enables us to directly measure the variation \h1\ abundance in these transition clouds. The timescale of the clouds had been examined through an analytical model. Our main results are the following. 

\begin{description}
\item{1.}
We identified a striking ``ring" of enhanced \h1\ abundance, which is the first time such a structure has been seen in a molecular cloud. It closely resembles the ``onion" shell description of a forming molecular cloud (Figure~\ref{fig:distribution}). The displacement between peak abundance positions of \h1, CO, and dust indicates the ongoing H$_2$ formation. Particularly, we find an orderly spatial variation of \h1\ abundance between 2\% and 0.2\%. 

\item{2.}
We derived the \h1$-$H$_2$ evolutionary timescales for these isolated dark clouds to be $\sim$6~Myr and further examined the cloud evolution through a time-dependent chemical model. 

\item{3.}
Our results could test H$_2$ formation models, constrain galaxy evolution simulations, and may affect future consideration of chemical evolution of disks, all of which indicate that the parameters of galactic H$_2$ formation play a crucial role for both galaxy evolution and star formation.  
\end{description}

\acknowledgments
This work is supported by National Key R\&D Program of China grant No.\ 2017YFA0402600, the National Natural Science Foundation of China grant No.\ 11725313 and No.\ 11690024, and the International Partnership Program of Chinese Academy of Sciences grant No.\ 114A11KYSB20160008. The Arecibo Observatory is operated by SRI International under a cooperative agreement with the National Science Foundation (AST-1100968), and in alliance with Ana G.~M\'{e}ndez-Universidad Metropolitana, and the Universities Space Research Association. 
D.L.\ acknowledges support from ``CAS Interdisciplinary Innovation Team" program. 
This work was carried out in part by the Jet Propulsion Laboratory, which is operated by NASA through the California Institute of Technology. 
Z.Y.Z.\ acknowledges support from ERC in the form of the Advanced
Investigator Programme, 321302, COSMICISM.



\clearpage
\begin{thebibliography}{}
          
          
\bibitem[Bergin et al.(2004)]{2004Bergin} Bergin, E.~A., Hartmann, L.~W., Raymond, J.~C., \& Ballesteros-Paredes, J.\ 2004, \apj, 612, 921 


\bibitem[Bialy et al.\ (2017)]{2017Bialy} Bialy, S., Bihr, S., Beuther, H., 
Henning, T., \& Sternberg, A.\ 2017, \apj, 835, 126 


\bibitem[Br{\"u}nken et al.(2014)]{2014Brunken} Br{\"u}nken, S., 
Sipil{\"a}, O., Chambers, E.~T., et al.\ 2014, \nat, 516, 219 


\bibitem[Burgh et al. (2010)]{2010Burgh} Burgh, E.~B., France, K., \& Jenkins, E.~B.\ 2010, \apj, 708, 334 


\bibitem[Caselli et al.(2002)]{2002Caselli} Caselli, P., Benson, P.~J., Myers, P.~C., \& Tafalla, M.\ 2002, \apj, 572, 238 


\bibitem[Cazaux et al.(2005)]{2005Cazaux} Cazaux, S., Caselli, P., Tielens, 
A.~G.~G.~M., LeBourlot, J., 
\& Walmsley, M.\ 2005, Journal of Physics Conference Series, 6, 155 


\bibitem[Cazaux 
\& Tielens(2004)]{2004Cazaux} Cazaux, S., \& Tielens, A.~G.~G.~M.\ 2004, \apj, 604, 222 


\bibitem[Elmegreen(2000)]{2000Elmegreen} Elmegreen, B.~G.\ 2000, \apj, 530, 
277 


\bibitem[Fukui et al.(2014)]{2014Fukui} Fukui, Y., Okamoto, R., Kaji, R., 
et al.\ 2014, \apj, 796, 59 




\bibitem[Gibson et al.(2000)]{2000Gibson} Gibson, S.~J., Taylor, A.~R., Higgs, L.~A., \& Dewdney, P.~E.\ 2000, \apj, 540, 851 


\bibitem[Glover 
\& Mac Low(2007)]{2007Glover} Glover, S.~C.~O., \& Mac Low, M.-M.\ 2007, \apj, 659, 1317 


\bibitem[Gnedin 
\& Kravtsov(2011)]{2011Gnedin} Gnedin, N.~Y., \& Kravtsov, A.~V.\ 2011, \apj, 728, 88 




\bibitem[Goldsmith 
\& Li(2005)]{2005Goldsmith} Goldsmith, P.~F., \& Li, D.\ 2005, \apj, 622, 938 

\bibitem[G{\"u}ver 
\& {\"O}zel(2009)]{2009Guver} G{\"u}ver, T., \& {\"O}zel, F.\ 2009, \mnras, 400, 2050 



\bibitem[Heeschen(1955)]{1955Heeschen} Heeschen, D.~S.\ 1955, \apj, 121, 
569 


\bibitem[Heiles 
\& Troland(2003)]{2003Heiles} Heiles, C., \& Troland, T.~H.\ 2003, \apj, 586, 1067 


\bibitem[Hincelin et al.(2013)]{2013Hincelin} Hincelin, U., Wakelam, V., 
Commercon, B., Hersant, F., \& Guilloteau, S.\ 2013, \apj, 775, 44 


\bibitem[Hincelin et al.(2011)]{2011Hincelin} Hincelin, U., Wakelam, V., 
Hersant, F., et al.\ 2011, \aap, 530, A61 


\bibitem[Hollenbach 
\& Salpeter(1971)]{1971Hollenbach} Hollenbach, D., \& Salpeter, E.~E.\ 1971, \apj, 163, 155 

\bibitem[Kaiser(2002)]{2002Kaiser}Kaiser, R.~I. \ 2002, Chemical Reviews, 102, 155



\bibitem[Knee 
\& Brunt(2001)]{2001Knee} Knee, L.~B.~G., \& Brunt, C.~M.\ 2001, \nat, 412, 308 


\bibitem[Kolpak et al.(2002)]{2002Kolpak} Kolpak, M.~A., Jackson, J.~M., Bania, T.~M., \& Dickey, J.~M.\ 2002, \apj, 578, 868 


\bibitem[Kramer et al.(1999)]{1999Kramer} Kramer, C., Alves, J., Lada, 
C.~J., et al.\ 1999, \aap, 342, 257 


\bibitem[Kr{\v c}o 
\& Goldsmith(2010)]{2010Krco} Kr{\v c}o, M., \& Goldsmith, P.~F.\ 2010, \apj, 724, 1402 


\bibitem[Lee et al.(2015)]{2015Lee} Lee, M.-Y., Stanimirovi{\'c}, S., 
Murray, C.~E., Heiles, C., \& Miller, J.\ 2015, \apj, 809, 56 


\bibitem[Li 
\& Goldsmith(2003)]{2003Li} Li, D., \& Goldsmith, P.~F.\ 2003, \apj, 585, 823 


\bibitem[Martin 
\& Barrett(1978)]{1978Martin} Martin, R.~N., \& Barrett, A.~H.\ 1978, \apjs, 36, 1 


\bibitem[Mathis et al.(1977)]{1977Mathis} Mathis, J.~S., Rumpl, W., \& Nordsieck, K.~H.\ 1977, \apj, 217, 425 


\bibitem[Minchin et al.(2007)]{2007Minchin} Minchin, R.~F., Auld, R., 
Davies, J.~I., et al.\ 2007, Galaxy Evolution across the Hubble Time, 235, 
227 


\bibitem[Peek et al.(2011a)]{2011Peeka} Peek, J.~E.~G., Heiles, C., Douglas, K.~A., et al.\ 2011a, \apjs, 194, 20 


\bibitem[Peek et al.(2011b)]{2011Peekb} Peek, J.~E.~G., Heiles, C., Peek, 
K.~M.~G., Meyer, D.~M., \& Lauroesch, J.~T.\ 2011b, \apj, 735, 129 



%
%
%
%




%
%
%







\bibitem[Sembach et al.(2001)]{2001Sembach} Sembach, K.~R., Howk, J.~C., Savage, B.~D., \& Shull, J.~M.\ 2001, \aj, 121, 992 


\bibitem[Semenov et al.(2010)]{2010Semenov} Semenov, D., Hersant, F., 
Wakelam, V., et al.\ 2010, \aap, 522, A42 


\bibitem[Shu(1977)]{1977Shu} Shu, F.~H.\ 1977, \apj, 214, 488 


\bibitem[Sofia et al. (2004)]{2004Sofia} Sofia, U.~J., Lauroesch, J.~T., Meyer, D.~M., \& Cartledge, S.~I.~B.\ 2004, \apj, 605, 272 




\bibitem[Sonnentrucker et al.(2007)]{2007Sonnentrucker} Sonnentrucker, P., Welty, D.~E., Thorburn, J.~A., \& York, D.~G.\ 2007, \apjs, 168, 58 


\bibitem[Spitzer(1946)]{1946Spitzer} Spitzer, L., Jr.\ 1946, Publications 
of the American Astronomical Society, 10, 144 

\bibitem[Spitzer(1978)]{1978Spitzer} Spitzer, L.\ 1978, Physical processes in the interstellar medium, by Lyman Spitzer.~ New York Wiley-Interscience, 1978.~333 p., 

\bibitem[Sternberg et al.(2014)]{2014Sternberg} Sternberg, A., Le Petit, F., Roueff, E., \& Le Bourlot, J.\ 2014, \apj, 790, 10 


\bibitem[Tennyson(2005)]{2005Tennyson} Tennyson, J.\ 2005, Astronomical 
spectroscopy : an introduction to the atomic and molecular physics of 
astronomical spectra, by J.~Tennyson.~Imperial College Press advanced 
physics texts, vol.~2 London: Imperial College Press, 2005 ISBN 
9781860946790,  








\bibitem[Vidali et al.(2005)]{2005Vidali} Vidali, G., Roser, J., 
Manic{\'o}, G., et al.\ 2005, Journal of Physics Conference Series, 6, 36 


\bibitem[Wakelam et al.(2012)]{2012Wakelam} Wakelam, V., Herbst, E., 
Loison, J.-C., et al.\ 2012, \apjs, 199, 21 


\bibitem[Walsh et al.(2015)]{2015Walsh} Walsh, C., Nomura, H., \& van Dishoeck, E.\ 2015, \aap, 582, A88 


\bibitem[Weselak et al.(2009)]{2009Weselak} Weselak, T., Galazutdinov, G.~A., Beletsky, Y., \& Kre{\l}owski, J.\ 2009, \mnras, 400, 392 


\bibitem[Woodall et al.(2007)]{2007Woodall} Woodall, J., Ag{\'u}ndez, M., Markwick-Kemper, A.~J., \& Millar, T.~J.\ 2007, \aap, 466, 1197 


\bibitem[Zemp et al.(2012)]{2012Zemp} Zemp, M., Gnedin, O.~Y., Gnedin, N.~Y., \& Kravtsov, A.~V.\ 2012, \apj, 748, 54 

\end{thebibliography}
\end{document}